\documentclass{IEEEoj}
\usepackage{cite}
\usepackage{amsmath,amssymb,amsfonts}
\usepackage{algorithmic}
\usepackage{graphicx,color}
\usepackage{textcomp}
\usepackage{subcaption}
\usepackage{float}
\usepackage{placeins}
\usepackage{hyperref}
\usepackage{url}
\usepackage{xurl}
\usepackage[font=footnotesize]{caption}
\usepackage{tikz}
\usetikzlibrary{arrows.meta, positioning}
\definecolor{navyblue}{RGB}{0,51,102}
\definecolor{navyfill}{RGB}{230,236,242}
\def\BibTeX{{\rm B\kern-.05em{\sc i\kern-.025em b}\kern-.08em
    T\kern-.1667em\lower.7ex\hbox{E}\kern-.125emX}}
\AtBeginDocument{\definecolor{ojcolor}{cmyk}{0.93,0.59,0.15,0.02}}
\def\OJlogo{\vspace{-14pt}\includegraphics[height=28pt]{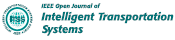}}
\begin{document}
\receiveddate{XX Month, XXXX}
\reviseddate{XX Month, XXXX}
\accepteddate{XX Month, XXXX}
\publisheddate{XX Month, XXXX}
\currentdate{XX Month, XXXX}
\doiinfo{OJITS.2026.3739219}

\title{Reinforcement Learning–Based Traffic Signal Control for IoT-Enabled Intersections}

\author{YOUSEF ALSAQABI, MEMBER, IEEE}
\affil{Department of Electrical Engineering, Kuwait University, Kuwait}
\corresp{yousef.alsaqabi@ku.edu.kw}
\markboth{Reinforcement Learning–Based Traffic Signal Control for IoT-Enabled Intersections}{AlSaqabi}

\begin{abstract}
Urban traffic congestion remains a persistent challenge in car-dependent cities, imposing significant economic and societal costs. Traffic signal systems are increasingly deployed as networked cyber-physical components within smart-city infrastructures, where distributed sensing and edge intelligence enable adaptive traffic management. This paper investigates reinforcement learning (RL) as an edge-intelligent approach for adaptive traffic signal operation at a signalized urban intersection in Kuwait. A Proximal Policy Optimization (PPO)-based controller is developed to dynamically allocate green-phase durations using locally observed traffic states, without relying on future demand information or centralized coordination. The controller is evaluated in a realistic simulation environment informed by real-world hourly traffic volume data from Kuwait, and is compared against both conventional fixed-time control and a vehicle-actuated controller representing the current state of practice, using average vehicle delay, queue length, and emissions as performance metrics. Under nominal conditions, the proposed controller reduces average vehicle delay by 46\% relative to fixed-time control and 34\% relative to actuated control, while also lowering per-vehicle CO\textsubscript{2} emissions by approximately 23\%. These performance gains persist under demand perturbations of $\pm15\%$, generalize from weekday to weekend traffic patterns, and are corroborated by a reward function ablation; low variance across five random seeds confirms their statistical reliability. These findings demonstrate the practicality of learning-based edge traffic signal control as a building block for IoT-enabled smart-city transportation systems, and as a deployable precursor toward fully connected, Internet of Vehicles (IoV)-based urban mobility.
\end{abstract}

\begin{IEEEkeywords}
Reinforcement Learning, Traffic Signal Control, Internet of Things, Internet of Vehicles, Smart Cities
\end{IEEEkeywords}


\maketitle

\section{INTRODUCTION}
\label{intro}
\subsection*{Urban Traffic Congestion: A Global and Kuwaiti Challenge}
Urban traffic congestion is a persistent global issue that contributes to travel delays, energy inefficiency, and increased air pollution. Its economic costs are substantial: U.S. cities alone experienced an estimated 3.6 billion vehicle hours of delay and \$67.5 billion in lost productivity annually in the early 2000s, which is equivalent to roughly 0.7\% of US GDP~\cite{texas2001}. More recent studies reported that congestion cost U.S. drivers \$166 billion in 2017, with commuters in major cities like Los Angeles losing up to 119 hours annually in traffic~\cite{texas2019}. Globally, INRIX estimated that urban congestion resulted in \$87 billion in annual losses across U.S. commuters in 2018 alone, with an average of \$1,348 per driver~\cite{inrix2018}.

In Kuwait, these challenges are amplified by high car dependency and a signal-heavy centralized road infrastructure. The country ranks among the world's highest in vehicle ownership, with over 2.67 million registered vehicles for a population of approximately 4.8 million: a ratio exceeding 560 vehicles per 1,000 people~\cite{ceic2025_kuwait, csb, helgi_kuwait_cars2019}. Private cars dominate daily mobility: more than 90\% of commuting trips are made using private vehicles, while public transit options remain extremely limited~\cite{ghadanfar}. This has placed considerable strain on Kuwait’s urban road network, with many major intersections and corridors experiencing severe bottlenecks and prolonged delays during peak traffic hours. Despite the objectives outlined in Kuwait Vision 2035~\cite{kuwait2035} to modernize infrastructure and promote smart mobility, deployment of intelligent traffic signal systems remains limited. In recent years, urban traffic signal systems have increasingly been viewed as networked cyber-physical infrastructure components within smart cities, where distributed sensing, communication, and edge intelligence jointly enable adaptive traffic management.

\subsection*{Limitations of Traditional Traffic Signal Control}
Conventional traffic signal control methods, such as fixed-time and actuated systems, dominate the current landscape in Kuwait and many similar regions. Fixed-time controllers rely on preset signal phase durations, often derived from outdated traffic surveys or simplistic demand assumptions. Actuated systems offer limited adaptability by responding to immediate traffic presence via detectors, but still operate based on fixed rules without contextual awareness or foresight. Both methods are inherently rule-based and fail to account for fluctuating traffic patterns, such as time-of-day variation, special events, or irregular traffic congestion. These approaches struggle in non-stationary traffic environments, resulting in inefficient signal timing, unnecessary waiting, and increased travel times. Given the diverse and evolving traffic behaviors across Kuwaiti intersections, a more adaptive and intelligent solution is needed.

Research on traffic modeling in Kuwait City shows that incorporating intersection-level delay functions into traffic assignment models improves prediction accuracy by 4–13\%, compared to traditional link-only approaches~\cite{alhaidar2016}. This suggests that standard control logic underrepresents the complexity of intersection-level congestion. Further simulation studies applying congestion pricing have demonstrated that better traffic and signal management can reduce total delay, fuel consumption, and overall user cost~\cite{alrukaibi}, reinforcing the importance of dynamic, adaptive control systems in Kuwait’s urban context.

Recent advances in artificial intelligence, particularly in reinforcement learning (RL), offer a promising path toward data-driven and context-aware traffic signal control within smart-city and Internet of Things (IoT)–enabled transportation infrastructures~\cite{RL_SLR}. In such settings, RL agents operate as edge-intelligent controllers that learn optimal decision policies through continuous interaction with sensed traffic states, potentially obtained via detectors, cameras, or vehicle-to-infrastructure communications, and adapt their actions based on observed system performance~\cite{RL}. This paradigm enables adaptive control under dynamic and uncertain traffic conditions without reliance on explicit traffic models or centralized coordination.

\subsection*{Motivation and Research Gaps}
Despite growing interest in data-driven traffic control systems, the majority of RL studies in this domain have been conducted in environments that are far removed from the practical realities of countries like Kuwait. Existing research tends to focus on idealized grid networks, synthetic traffic flows, or cities with extensive sensing infrastructure and high-resolution datasets. These assumptions rarely hold in Gulf countries, where intersections vary widely in topology, public data is scarce, and real-time sensor coverage is limited. Additionally, most RL-based studies evaluate performance under a narrow set of conditions without testing the system’s generalization to unseen patterns, sensitivity to data noise, or robustness to variability in traffic volume.

In the context of Kuwait, this lack of regional and methodological coverage creates a clear research opportunity, and the significance of the setting lies not in its geography but in the distinct operating regime it represents. This regime differs from the environments in which RL-based signal control has predominantly been evaluated in four respects. First, the demand structure is shaped by the extreme car dependency documented above: with private vehicles carrying nearly all commuting trips and minimal transit alternatives, intersections absorb almost the entirety of urban mobility, producing sustained high volumes rather than the transit-moderated demand profiles of the cities underlying most existing evaluations. Second, the weekend demand pattern itself is distinctive: weekend days in Kuwait, and in the region more broadly, are shaped by religious and social practices, most prominently midday congregational prayers on Friday, producing weekend demand profiles whose temporal structure differs from those of the regions underlying most existing evaluations, so cross-day transferability must be demonstrated on these patterns rather than assumed from existing studies. Third, the sensing environment is genuinely data-scarce: traffic authorities operate fixed roadway sensors producing aggregate counts, without the trajectory, probe, or camera data that many RL approaches presuppose, and whether learning-based control is feasible under this sensing constraint is precisely what a deployment-relevant evaluation must establish. Fourth, adaptive signal control is virtually absent from the region's intersections, so the operationally meaningful question for traffic authorities is the improvement attainable over the fixed-time and actuated control deployed today. Under these conditions of sustained demand, coarse sensing, a distinct temporal structure, and a conventional-control baseline, the transferability of conclusions from data-rich, transit-oriented settings cannot be assumed, and the present study evaluates RL-based signal control under exactly this regime.

From an IoT and Internet of Vehicles (IoV) perspective, these limitations further highlight the dependence of data-driven traffic control on large-scale vehicular data, which in practice is often constrained by privacy concerns, data ownership, and incentive mechanisms in vehicular networks. Prior work has highlighted these challenges and proposed decentralized, privacy-preserving frameworks to enable vehicular data aggregation under realistic participation constraints~\cite{alsaqabi_privacy, alsaqabi2026}.

These connectivity constraints directly motivate the sensing approach adopted in this work. It is therefore important to distinguish between the sensing modalities such systems may employ. The controller developed here relies on IoT-based infrastructure sensing: aggregate vehicle counts from fixed roadway sensors. This sensing modality is deployable in Kuwait today and representative of the data accessible to current traffic authorities. This is distinct from vehicle-to-infrastructure (V2X) sensing, which depends on high penetration of connected vehicles and mature communication infrastructure not yet widely available in the region. Because the controller relies on aggregate counts from fixed infrastructure sensors rather than data transmitted by individual connected vehicles, its operation is independent of connected-vehicle penetration rate, a practical advantage in regions where vehicle connectivity remains limited. We therefore position IoT-infrastructure-based RL control as a necessary and immediately deployable precursor to fully connected, IoV-based control, with the integration of connected-vehicle data and the modeling of communication-layer effects such as latency and packet loss identified as directions for future work.

\subsection*{Contributions}

This study addresses two key gaps in the existing literature:  
(1) a \textbf{contextual gap}, arising from the lack of RL-based traffic signal control studies in Gulf and MENA urban environments such as Kuwait, and  
(2) an \textbf{evaluation gap}, reflecting limited understanding of how learning-based edge controllers perform under realistic, fluctuating, and imperfect traffic sensing conditions.

Within the broader context of smart-city IoT and IoV systems, this work makes the following contributions:

\begin{itemize}
    \item We design a reinforcement learning--based edge controller for adaptive traffic signal operation, modeling the signalized intersection as an intelligent IoT node within a smart transportation infrastructure.

    \item We integrate the controller with a simulation framework informed by real-world traffic measurements, enabling realistic evaluation under sensing-limited and non-ideal traffic conditions representative of Kuwait’s urban environment.

    \item We conduct a comprehensive experimental evaluation that benchmarks the controller against both fixed-time and vehicle-actuated baselines, examining robustness to demand uncertainty, cross-day generalization (weekday to weekend), reward function sensitivity, and environmental impact, providing insights into the reliability and transferability of learning-based signal control in networked urban systems.
    
\end{itemize}

%%%%%%%%%%%%%%%%%%%%%%%%%%%%%
The remainder of this paper is organized as follows. 
Section~\ref{sec:relatedwork} reviews related work on traffic signal control using both traditional approaches and reinforcement learning–based methods. 
Section~\ref{sec:methodology} describes the proposed methodology, including the problem formulation, PPO-based control framework, and simulation setup.
Section~\ref{sec:experiments} details the experimental design and evaluation metrics used in this study. 
Section~\ref{sec:results} presents and discusses the experimental results. 
Finally, Section~\ref{sec:conclusion} concludes the paper and highlights directions for future research.

\section{RELATED WORK}
\label{sec:relatedwork}
\subsection{Traditional Traffic Signal Control}

Traditional traffic signal control has developed over nearly a century, progressing from manual operation and fixed-cycle plans to increasingly automated and adaptive systems. Fixed-time controllers, first formalized in the mid-twentieth century, remain among the most prevalent worldwide. They operate using pre-timed signal plans that are designed offline based on historical or manually collected traffic data, and are effective only under stable demand patterns~\cite{gordon1996traffic,li2014survey}. While simple and inexpensive to deploy, their inability to respond to real-time fluctuations often leads to unnecessary delays and fuel wastage during off-peak or abnormal conditions~\cite{papageorgiou2003review}. 

Actuated controllers represented the next major evolution by integrating vehicle detection technologies such as inductive loops, magnetometers, and radar to adjust green times when demand is sensed. Despite offering localized adaptability, these systems still rely on deterministic rules, threshold-based parameters, and predefined minimum and maximum green limits~\cite{cress2024roadsideITS,li2014survey}. Cre{\ss}~\textit{et al.}~(2024) note that early intelligent transportation initiatives, including the European \textit{PROMETHEUS} project and the California \textit{PATH} program, utilized such rule-based occupancy detection but lacked predictive or learning-based mechanisms for adaptive decision-making~\cite{cress2024roadsideITS}. 

To address the shortcomings of isolated actuated controllers, fully adaptive and coordinated control systems such as SCOOT (Split Cycle Offset Optimization Technique) and SCATS (Sydney Coordinated Adaptive Traffic System) were developed in the late twentieth century. These systems continuously adjust signal timings using aggregated flow data from field sensors and can achieve up to 10--20\% improvements in travel time when properly calibrated~\cite{gordon1996traffic,papageorgiou2003review,stevanovic2010stochastic}. However, they are highly infrastructure-dependent, requiring extensive sensor networks, centralized computing, and recurring human supervision to maintain performance~\cite{cress2024roadsideITS}. As emphasized by Gordon~\textit{et al.}, most adaptive systems rely on heuristic optimization and linear control logic, making them unsuitable for rapidly changing or stochastic conditions. 

Even with recent advances in coordinated and multi-agent signal control, such as heuristic-based cooperative systems and fuzzy-logic controllers~\cite{srinivasan2006multiagent,teodorovic2000fuzzy}, traditional approaches continue to operate under rule-based paradigms. Their central decision policies are predefined rather than learned, and thus they cannot adapt autonomously to new patterns of congestion or incidents. From an infrastructure standpoint, classical intelligent transportation systems rely heavily on roadside sensing, including loop detectors, cameras, and radar units, for flow measurement and queue estimation~\cite{cress2024roadsideITS,ogie2017smartinfra,liu2007survey}. These sensors underpin today’s cooperative-ITS (C-ITS) networks, enabling vehicle–infrastructure communication via roadside units (RSUs), but the control logic within these systems remains deterministic rather than data-driven~\cite{cress2024roadsideITS,liu2007survey}. 

In summary, traditional and rule-based signal control methods, covering fixed-time, actuated, and adaptive systems, have formed the backbone of modern traffic management, but remain constrained by static decision rules and dependence on centralized, sensor-intensive infrastructure. Such controllers cannot adapt to new or unpredictable traffic conditions, nor can they improve performance through experience, limiting their effectiveness in highly variable urban settings~\cite{papageorgiou2003review,liu2007survey,li2014survey}.

The work presented in this paper introduces a reinforcement learning–based signal controller designed to automatically adapt to changing traffic patterns using real-world data from Kuwait. The proposed system is systematically evaluated for its generalization across different times of day and between weekdays and weekends, its robustness to fluctuations in traffic volume, and the influence of reward function design. Through these evaluations, the study addresses several limitations of conventional control methods and demonstrates the potential of data-driven learning approaches for more responsive urban traffic management.

\begin{figure}[H]
   \centering
    \graphicspath{{./Figs/}}
    \includegraphics[scale=0.5]{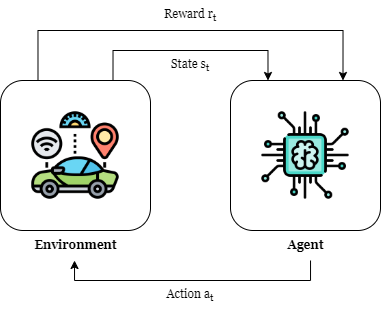}
    \caption{Reinforcement learning control loop.~\cite{alsaqabi}}
    \label{fig:loop}
\end{figure}

\subsection{Reinforcement Learning for Traffic Signal Control}

Reinforcement Learning (RL) is defined as a method by which an agent interacts with its environment to learn optimal behaviors without prior knowledge~\cite{bae}. The objective of RL is to maximize cumulative numerical rewards or equivalently minimize penalties through repeated interaction, where feedback received after each action enables the agent to iteratively refine its control policy~\cite{RL}. As illustrated in Figure~\ref{fig:loop}, at each time step $t$ the agent observes the current state $s_t$, selects an action $a_t$ according to its policy, and receives a reward $r_t$, which together drive policy improvement over time.

Within intelligent transportation systems, this agent–environment formulation has been adopted at multiple control layers, from vehicle-level decision-making to infrastructure-level traffic management. Reinforcement learning has been explored, for example, for data-aware autonomous vehicle path planning under dynamic traffic conditions~\cite{alsaqabi}. At the infrastructure level, RL has emerged as a prominent alternative to model-driven signal optimization, as it learns control policies directly from observed traffic without assuming flow models or stationarity~\cite{RL_SLR, RLTechniques, Saadi2025ASO}. The signal controller is modeled as an agent in a Markov decision process: it observes the traffic state, selects a control action (e.g., extending or changing a phase), and receives a reward encoding objectives such as delay, throughput, or queue reduction~\cite{RL}. This data-driven view is attractive where traffic is highly non-stationary and classical fixed-time or actuated strategies struggle to adapt in real time.

\subsubsection{Taxonomy of Current RL Approaches}
Existing work in RL-based traffic signal control can be grouped into four main families according to their algorithmic structure and deployment focus~\cite{RL_SLR, RLTechniques, chu2020marl, devailly2021ig}.

\textit{Value-based methods} (Q-learning, DQN, Double-DQN) estimate expected return per action and are common for single-intersection problems; Li~\textit{et al.}~\cite{li2016traffic} established one of the first DRL formulations for adaptive signal timing~\cite{Q, RLTechniques}.

\textit{Policy-gradient and actor–critic methods} (A3C, DDPG, PPO) directly learn a parameterized policy, offering smoother control in continuous action spaces~\cite{peters, bhasin}; Mao~\textit{et al.}~\cite{mao2022mastering} applied soft actor–critic to arterial coordination.

\textit{Graph and attention-based multi-agent methods} treat each intersection as a node exchanging embeddings with neighbors for city-scale cooperation~\cite{zhou2024cooperative}; CoLight~\cite{wei2019colight} uses graph attention to weight neighbor information, and IG-RL~\cite{devailly2021ig} uses an inductive representation that generalizes to unseen topologies without retraining.

\textit{Meta- and transfer-learning} approaches such as MetaLight accelerate adaptation to unseen traffic by sharing knowledge across scenarios.

While all these categories share the goal of adaptive timing, they differ in sample efficiency, scalability, and interpretability. Surveys note that value-based methods remain data-hungry and unstable; policy-gradient methods scale better but require careful tuning; and graph/meta-learning improve coordination but add complexity and overhead~\cite{RL_SLR, RLTechniques, chu2020marl, devailly2021ig}.

\subsubsection{Applications and Extensions of RL in Traffic Signal Control}

Recent work extends RL beyond single intersections toward cooperative network-level control, where multi-agent frameworks let each signal act as an agent coordinating with neighbors to optimize global flow~\cite{chu2020marl, wei2019colight, devailly2021ig}. More recently, Othman et al.~\cite{Othman} proposed a decentralized multi-agent RL controller that incorporates multimodality and partial observability, reflecting the continued movement toward scalable, decentralized coordination. A second extension integrates connected and automated vehicles (CAVs), exploiting vehicle trajectory and speed data through vehicle-to-infrastructure communication. Feng~\textit{et al.}~\cite{feng2015real} showed substantial delay reduction using connected-vehicle data, and frameworks such as CVLight~\cite{mo2022cvlight} and CoTV~\cite{dai2024marp} demonstrated cooperative signal–vehicle control. However, their reliance on dense communication and high CAV penetration limits deployment in data-scarce environments.

\subsubsection{Open Challenges and Motivation}
Across recent surveys and benchmark evaluations, three recurring research gaps have been consistently identified:  
\begin{itemize}
    \item \textbf{Limited realism in experimental settings:} Most RL-based traffic signal controllers are evaluated on homogeneous or idealized networks that lack stochastic demand, measurement noise, or heterogeneous driver behavior, reducing their real-world applicability~\cite{RLTechniques, RL_SLR}.  

    \item \textbf{Insufficient testing of temporal and demand variability:} Few studies explicitly assess how learned policies generalize across time-of-day fluctuations, weekday–weekend traffic differences, or systematic demand shifts, leaving robustness under real traffic dynamics largely unexplored~\cite{zhou2024cooperative, mao2022mastering}. Recent efforts have begun to address this; for example, Shi~\textit{et al.}~\cite{Shi2024} study the robustness and generalizability of large-scale RL signal control under sensor failures and transfer to unseen network topologies, and Devailly~\textit{et al.}~\cite{devailly2024mujam} extend generalization to unseen road networks and controller constraints through model-based graph reinforcement learning. However, such work focuses on city-scale network transfer rather than robustness and cross-day generalization on real single-intersection demand, which is the focus of this study. 

    \item \textbf{High reliance on connected-vehicle penetration:} RL frameworks designed for connected or mixed-autonomy environments commonly assume high levels of vehicle connectivity, which limits their transferability to regions with emerging or partial CAV adoption~\cite{feng2015real, mo2022cvlight, tajalli2021traffic}.
\end{itemize}

To address these challenges, this study develops and evaluates a reinforcement learning–based traffic signal controller using real hourly traffic volume data from Kuwait. The proposed framework systematically evaluates baseline performance against both fixed-time and vehicle-actuated control, tests robustness to moderate traffic demand perturbations, and examines generalizations across distinct weekday and weekend traffic regimes. Additionally, reward function ablation experiments evaluate the impact of individual reward components on learning stability and control performance. Collectively, these experiments characterize the reliability and transferability of learning-based signal control under realistic, data scarce conditions.

%%%%%%%%%%%%%%%%%%%%%%%%%%%%%%%%%%%%%%%%%%%%%%%%%%%%%%
\subsection{Traffic Management in the Gulf and MENA Region}

Research applying machine learning or intelligent control to traffic management in the Gulf and broader MENA region remains limited compared with global literature, with only a handful of studies published in high-impact venues. One of the earliest regional efforts is by Ghazal \textit{et al.}~\cite{ghazal2016smart}, who designed a sensor-driven smart traffic light system using infrared and ultrasonic detectors integrated with microcontrollers to dynamically allocate green time based on lane occupancy. Their prototype demonstrated significant delay reduction in field trials, establishing one of the first intelligent signal control frameworks developed in the Arab region. Building on this foundation, Natafgi \textit{et al.}~\cite{natafgi2018smart} proposed a machine-learning-based traffic light controller that employed supervised learning models to adjust phase timing in real time based on camera-detected vehicle density. Implemented on an embedded platform and experimentally validated, their approach achieved improved throughput compared with fixed-time control, marking a key regional step toward data-driven adaptation.

Berbar \textit{et al.}~\cite{berbar2022reinforcement} propose a reinforcement-learning-based signal controller for signalized intersections with connected and autonomous vehicle (CAV) platoons. The framework uses a two-agent architecture: one RL agent adjusts the approach speeds of incoming vehicle platoons to reduce fuel consumption, while a second RL agent controls the signal phases to minimize intersection delay. In Kuwait, Almomany \textit{et al.}~\cite{almomany2025real} recently introduced a reinforcement-learning-enhanced urban traffic signal optimization framework motivated by conditions in Kuwait City. Their study systematically compared multiple control strategies (fixed-time, max-pressure, delay-based optimization, and a reinforcement learning controller) under a range of traffic demand scenarios using microscopic simulation. In Qatar, Shaaban and Ghanim~\cite{shaaban2018evaluation} used microsimulation to study Transit Signal Priority (TSP) along a major arterial in central Doha. Their VISSIM-based model tested priority strategies that extend green or truncate red for buses and showed that targeted signal priority can cut transit travel time by over 40\% during peak periods, with minimal negative impact on general traffic. Although this work is not reinforcement learning, it is important regional evidence that policy-driven, algorithmic priority control can materially change corridor performance in a Gulf city where private car dependence is very high and bus ridership is still developing.

In summary, prior Gulf and MENA studies demonstrate early progress toward intelligent traffic control but remain limited, which motivates the present work. Earlier regional efforts relied on rule-based or supervised sensing systems that assume idealized detection and connectivity~\cite{ghazal2016smart,natafgi2018smart}, while CAV-oriented frameworks~\cite{berbar2022reinforcement} presuppose connected-vehicle penetration not yet present in the region. Most directly related is the recent work of Almomany~\textit{et al.}~\cite{almomany2025real}, which also studies reinforcement learning for traffic signal control in a Kuwait context. That study, however, differs from ours in several fundamental respects. First, its experiments use synthetic demand generated via random trip sampling on a generic, automatically generated grid network, whereas the present work is driven by real hourly traffic volume data collected from fixed sensors at an actual signalized intersection on Kuwait's Second Ring Road, with geometry derived from real map data. Second, reinforcement learning in their framework is one of several control strategies benchmarked in service of a hardware-acceleration contribution (FPGA-based execution), whereas our work is a dedicated study of a learning-based controller and its behavior. Third, and most importantly, we systematically evaluate the learned policy along dimensions their study does not address: robustness to demand perturbation, generalization from weekday to weekend traffic regimes, and sensitivity to reward function design. The distinction matters because it is precisely the regime characteristics
described in Section~\ref{intro}, which are sustained car-dependent demand, coarse infrastructure sensing, and a distinct weekly demand pattern, that a synthetic grid-network evaluation cannot capture. To our knowledge, this constitutes the first empirical evaluation of reinforcement learning–based adaptive signal control on real traffic data from a Gulf urban intersection, with explicit testing of policy robustness and transferability under realistic, data-scarce conditions.

\section{Methodology}
\label{sec:methodology}
\subsection{Traffic Control Environment}
\begin{figure}[t]
    \centering
    \includegraphics[width=0.7\columnwidth]{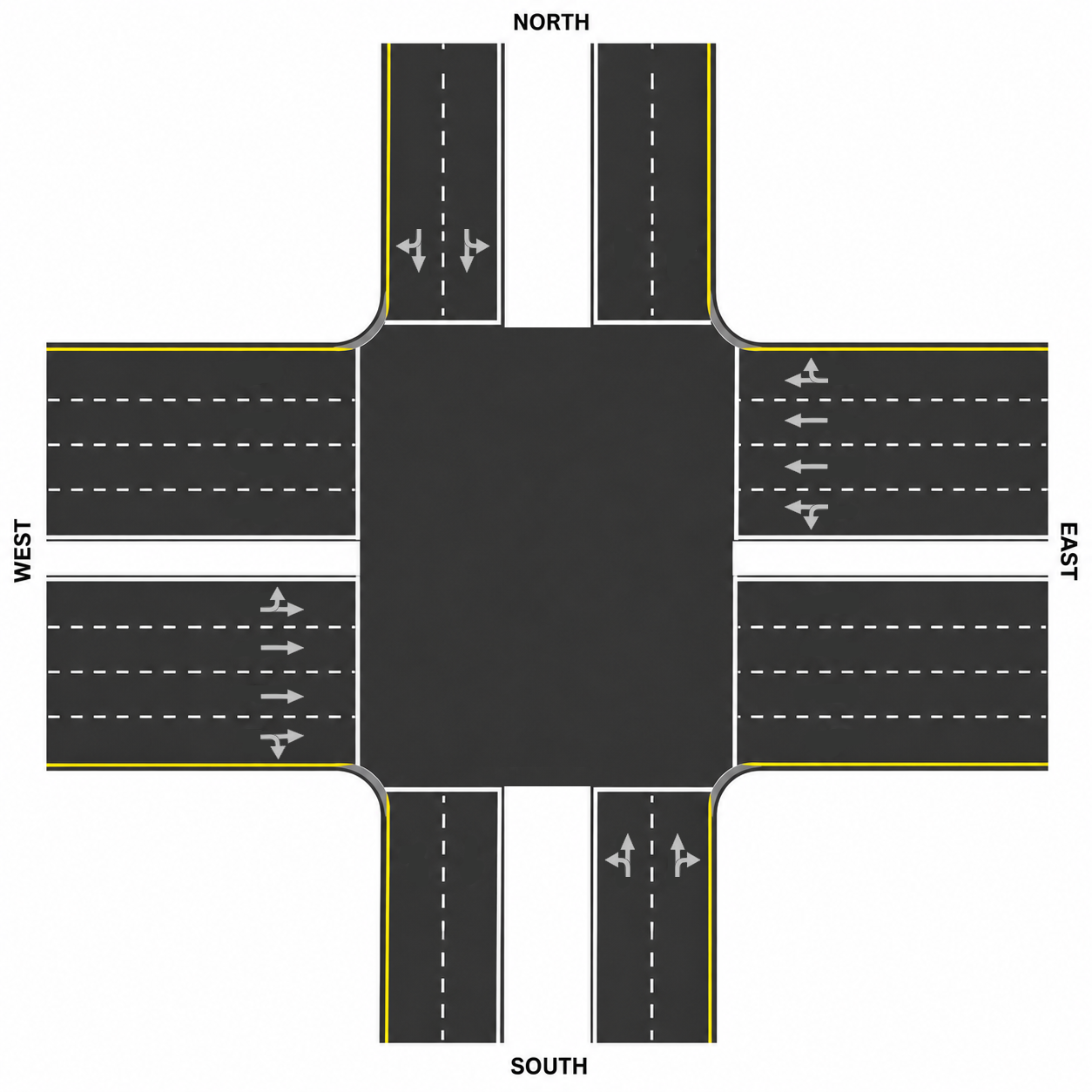}
    \caption{Layout of the studied four-leg signalized intersection on Kuwait's Second Ring Road, showing the lane configuration and permitted turning movements for each approach. The east–west arterial carries four lanes per direction and the north–south road two lanes per direction. Geometry is based on the OpenStreetMap-derived SUMO network.}
    \label{fig:intersection}
\end{figure}

We model traffic signal control using a microscopic simulation environment implemented in the Simulation of Urban MObility (SUMO)~\cite{Sumo} platform. The environment represents a real, signalized four-leg intersection located on Kuwait’s Second Ring Road, one of the city’s primary roads. The intersection geometry and lane connectivity are derived from OpenStreetMap~\cite{OpenStreetMap}, ensuring layout accuracy while enabling reproducible simulation-based experimentation. Microscopic simulation is adopted to capture vehicle-level interactions, including car-following, lane-changing, and signal compliance, which are essential for accurately modeling congestion and delay at signalized intersections. The layout of the studied intersection is illustrated in Fig.~\ref{fig:intersection}.

\subsubsection{Intersection and Signal Configuration}
The modeled intersection consists of multiple incoming and outgoing approaches with dedicated lanes for straight and turning movements, all controlled by a single traffic signal. The signal controller follows a predefined cyclic phase program consisting of four green phases, each followed by a corresponding clearance (yellow) phase. The phases serve, in sequence: (1) east–west through and right movements, (2) east–west left turns, (3) north–south through and right movements, and (4) north–south left turns. The phase sequence and signal states are fixed to enforce safety constraints and maintain consistency across control strategies.

While the phase order remains unchanged, the duration of green phases is adaptively controlled, allowing the signal controller to allocate green time based on observed traffic conditions. Clearance intervals are enforced according to the predefined signal program and are not subject to learning. This design isolates the learning task to green-time allocation while preserving realistic signal operation.

\subsubsection{Traffic Demand Modeling}
Traffic demand is modeled using real traffic volume data obtained from Kuwait’s Ministry of Interior~\cite{KuwaitMOITrafficData}, collected via fixed underground roadway sensors installed at the studied intersection. These sensors provide aggregate vehicle counts by detecting passing vehicles and are commonly used by traffic authorities for monitoring and signal planning purposes.

The collected counts are processed into time-varying flow rates that capture typical urban traffic patterns, including variations across days. Traffic demand is represented as piecewise-constant vehicle flows defined over consecutive short intervals, enabling fine-grained temporal variation while maintaining computational tractability. This approach reflects realistic fluctuations in traffic intensity while avoiding reliance on fine-grained vehicle trajectory data.

Each flow corresponds to a specific origin–destination movement through the intersection, thereby implicitly encoding turning behavior. Vehicle arrivals follow a stochastic insertion process consistent with the specified average flow rates, introducing realistic variability in arrival times while preserving the underlying demand profile. All vehicles are modeled as passenger cars with homogeneous physical and behavioral characteristics. Using infrastructure-based vehicle counts ensures that the simulated demand is grounded in measurements obtainable by existing traffic management systems.

\subsubsection{Environment Observations}
At each control decision epoch, the environment provides the signal controller with an observation vector composed of aggregated lane-level traffic measurements and signal state information. For each lane controlled by the signal, two congestion indicators are included: (i) queue length, measured as the number of halted vehicles, and (ii) cumulative waiting time.

The controller also observes the currently active green phase and the elapsed time spent in that phase. The resulting state representation is low-dimensional, interpretable, and relies solely on information that would be available in a practical deployment.

\subsubsection{Simulation Timing and Episode Execution}
The simulation advances in discrete one-second time steps. Control decisions are applied at the beginning of each green phase, after which the simulator advances for the selected green duration, followed by the associated clearance interval. Traffic dynamics evolve continuously during this period, and performance metrics are accumulated internally.

Each simulation episode corresponds to a fixed control horizon representing a full 24-hour cycle of simulated time. An episode terminates when the predefined horizon is reached or when no further vehicles remain in the network. This episodic formulation enables consistent evaluation across varying traffic conditions.

\subsubsection{Scope and Assumptions}
The environment models a single isolated intersection without coordination or communication with neighboring signals. No vehicle-to-infrastructure communication or future demand knowledge is assumed at execution time, and all control decisions are made by a pre-trained policy based solely on currently observed traffic conditions and signal state. This setup reflects a realistic deployment scenario for adaptive signal control at standalone urban intersections and provides a controlled foundation for future extensions.

\subsection{Reinforcement Learning Framework}

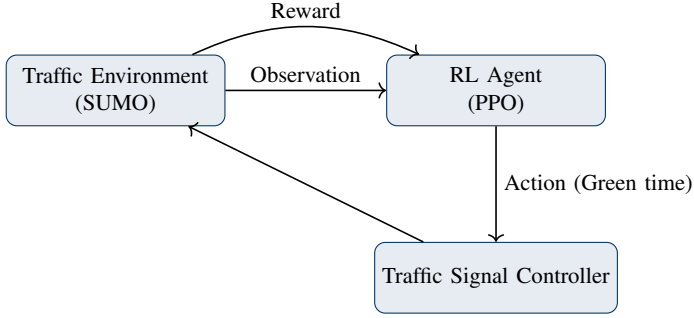
\begin{figure}[t]
\centering
\definecolor{navyblue}{RGB}{0,51,102}
\definecolor{navyfill}{RGB}{230,236,242}

\begin{tikzpicture}[
    scale=0.85,
    transform shape,
    block/.style={
        draw=navyblue,
        fill=navyfill,
        rectangle,
        rounded corners,
        align=center,
        minimum width=3.4cm,
        minimum height=1.1cm
    },
    arrow/.style={->, semithick}
]

\node[block] (env) {Traffic Environment\\(SUMO)};
\node[block, right=2.5cm of env] (agent) {RL Agent\\(PPO)};
\node[block, below=1.8cm of agent] (ctrl) {Traffic Signal Controller};

% Observation (top horizontal)
\draw[arrow] (env) -- node[above]{Observation} (agent);
% Reward (separate arrow so it can't sit under Observation)
\draw[arrow] (env) to[bend left=25] node[above]{Reward} (agent);

% Action (vertical)
\draw[arrow] (agent) -- node[right]{Action (Green time)} (ctrl);

% Environment transition (diagonal)
\draw[arrow] (ctrl) -- (env);

\end{tikzpicture}
\caption{Closed-loop interaction between the reinforcement learning agent and the traffic simulation environment.}
\label{fig:rl_loop}
\end{figure}

\subsubsection{Reinforcement Learning Formulation}
Adaptive traffic signal control is formulated as a Markov Decision Process (MDP)~\cite{RL}, in which the traffic signal controller acts as an agent interacting with the signalized intersection environment described in the previous section. At each decision epoch, the agent observes the current traffic state and selects a control action that determines the duration of the active green phase. State observations consist of aggregated lane-level congestion indicators and signal state information, ensuring that control decisions rely solely on measurable quantities available in practical deployments.

The action space is continuous and corresponds to selecting the duration of the currently active green phase within predefined minimum and maximum bounds. These bounds are imposed to ensure safe and realistic signal operation and to prevent excessively short or long green intervals. The phase sequence itself is fixed and follows standard traffic signal safety constraints; only the allocation of green time is subject to control. This formulation isolates the learning problem to green-time optimization while preserving realistic signal behavior and regulatory compliance.

The reward function is defined as a weighted combination of throughput and congestion-related penalties computed at each control decision interval. Specifically, the instantaneous reward $r_t$ is given by

\begin{equation}
r_t = 
w_a \, N_t
- w_q \, Q_t
- w_w \, \frac{W_t}{\kappa},
\label{eq:reward}
\end{equation}

where $N_t$ denotes the number of vehicles that successfully traverse the intersection during interval $t$, $Q_t$ represents the total number of queued vehicles across all controlled lanes, and $W_t$ is the cumulative waiting time of vehicles. The constant $\kappa$ is a scaling factor used to normalize the magnitude of the waiting-time term, while the non-negative coefficients $w_a$, $w_q$, and $w_w$ control the relative importance of throughput maximization versus congestion mitigation.

To ensure stable learning behavior, congestion-related penalty terms are normalized by the number of controlled lanes, and the resulting reward is clipped within predefined bounds to prevent extreme values that could destabilize policy optimization. Overall, this formulation encourages efficient vehicle discharge while discouraging excessive queue buildup and prolonged waiting, using only aggregate measurements obtainable from standard traffic monitoring infrastructure.

\subsubsection{Learning Algorithm and Training Procedure}

\begin{table}[!t]
\centering
\caption{PPO hyperparameters and training configuration.}
\label{tab:hyperparams}
\footnotesize
\setlength{\tabcolsep}{4pt}
\begin{tabular}{ll}
\hline
\textbf{Parameter} & \textbf{Value} \\
\hline
\multicolumn{2}{l}{\textit{Network Architecture}} \\
Policy / value networks & MLP, [128, 128] \\
Activation function & Tanh \\
\hline
\multicolumn{2}{l}{\textit{PPO Algorithm}} \\
Learning rate & $3 \times 10^{-4}$ \\
Discount factor $\gamma$ & 0.995 \\
GAE parameter $\lambda$ & 0.97 \\
Clipping parameter $\epsilon$ & 0.1 \\
Entropy coefficient & 0.01 \\
Value function coefficient & 0.5 \\
Max gradient norm & 0.5 \\
Rollout length & 2048 \\
Batch size & 1024 \\
Epochs per update & 10 \\
Total training timesteps & 500{,}000 \\
\hline
\multicolumn{2}{l}{\textit{Evaluation}} \\
Evaluation seeds & 5 (0--4) \\
\hline
\multicolumn{2}{l}{\textit{Reward Function (Eq.~\ref{eq:reward})}} \\
Throughput weight $w_a$ & 0.3 \\
Queue weight $w_q$ & 0.3 \\
Waiting weight $w_w$ & 0.3 \\
Scaling factor $\kappa$ & 10.0 \\
Reward clipping & $[-100, 100]$ \\
\hline
\multicolumn{2}{l}{\textit{Signal Timing Constraints}} \\
Minimum green time & 15 s \\
Maximum green time & 60 s \\
Yellow clearance & 3 s \\
\hline
\end{tabular}
\end{table}

To learn an effective control policy, we employ a policy-gradient-based reinforcement learning approach using Proximal Policy Optimization (PPO)~\cite{PPO}. PPO is an actor–critic method that performs stable policy updates through a clipped objective function, making it well-suited for control tasks under stochastic and time-varying traffic demand.~\cite{konda2003onactor} The policy and value functions are parameterized by feedforward neural networks and trained jointly.

Training is conducted offline using episodic interaction with the simulation environment. During training, the agent collects rollouts over fixed-length episodes corresponding to a full 24-hour cycle of simulated time and updates the policy parameters based on accumulated experience. Once training is complete, the learned policy is fixed and used for control without further online adaptation. During execution, all control actions are generated by the pre-trained policy using only the current traffic observations and signal state, without access to future demand information or external communication. This separation between offline learning and online execution ensures both stable training and realistic deployment assumptions.

The complete set of PPO hyperparameters, network architecture, and training configuration is summarized in Table~\ref{tab:hyperparams}. These settings were held fixed across all experiments. The reward weights shown correspond to the full reward formulation; the ablation study (Experiment~4) varies these weights, and the generalization study (Experiment~3) varies the training demand data.

\subsection{Simulation Validity and Sim-to-Real Considerations}

The credibility of simulation-based evaluation depends on how truly the simulation reflects real-world conditions. This subsection describes which aspects of the present study are grounded in real-world data, which rely on established simulation models, and which are approximated, in order to characterize the gap between simulation and field deployment.

Two components of the experimental setup are derived directly from real-world sources. First, the traffic demand is constructed from real hourly vehicle volume counts collected by fixed roadway sensors operated by the Kuwait Ministry of Interior at the studied intersection, rather than from synthetic or assumed flows. Second, the intersection geometry and lane connectivity are derived from OpenStreetMap data for an actual signalized intersection on Kuwait's Second Ring Road, preserving the real layout, approach structure, and turning connections. Consequently, both the demand the controller experiences and the physical network on which it operates are empirically grounded.

Vehicle-level dynamics are governed by SUMO's microscopic car-following and lane-changing models, which reproduce realistic speed–headway and flow–density relationships and capture the vehicle interactions that drive queue formation and dissipation at signalized intersections. Microscopic simulation of this kind is widely used as a credible surrogate for intersection-level traffic behavior. Nonetheless, several parameters are not independently calibrated against field measurements and are instead set to established simulation defaults. Specifically, saturation flow rates and discharge headways follow the simulator's default car-following parameterization rather than values measured at the site; turning-movement proportions are encoded implicitly through the origin–destination structure of the demand but are not independently validated against observed turning counts; and all vehicles are modeled as homogeneous passenger cars, omitting heavy-vehicle effects and heterogeneity in driver behavior. The model likewise assumes full signal compliance and does not represent phenomena such as red-light running or stop-line encroachment.

These approximations define the principal sources of sim-to-real uncertainty. While direct field validation, which includes comparing simulated queues, delays, and discharge rates against observed intersection behavior, would most rigorously establish transferability, such validation requires operational access to instrumented intersection performance data beyond the aggregate volume counts available for this study, and is subject to approval by the relevant traffic authority. We therefore identify field-level calibration and validation as an important direction for future work. In the interim, the robustness and generalization experiments reported in Section~\ref{sec:results} provide indirect evidence relevant to transfer: the $\pm 15\%$ demand-perturbation experiments assess sensitivity to errors in demand magnitude of the kind expected from imperfect real-world counts, while the weekday-to-weekend generalization experiments assess robustness to shifts in demand pattern not seen during training. The controller's ability to retain performance under both forms of distributional mismatch, without retraining, offers preliminary evidence that the learned policy does not overfit to its exact training distribution, which is a necessary condition for transfer beyond the simulator. These two tests jointly cover the two principal axes of demand uncertainty, which are magnitude and temporal pattern, but do not exhaust the robustness dimensions relevant to deployment: sensor failures, incidents, adverse weather, and special events lie outside the scope of the available data and remain directions for future evaluation.

\section{EXPERIMENTAL EVALUATION}
\label{sec:experiments}

This section describes the experimental setup and evaluation protocol used to assess the performance of the proposed reinforcement learning--based traffic signal controller. All experiments are conducted using the simulation environment described in Section~\ref{sec:methodology} under realistic traffic demand derived from real-world measurements. The objective is to evaluate the effectiveness, generalizability and robustness of the learned policy under varying traffic conditions. All reported percentage improvements are computed over the full 24-hour evaluation horizon.

\subsection{Experimental Setup, Assumptions, and Baselines}

The proposed controller is trained using Proximal Policy Optimization (PPO) within the SUMO-based traffic environment. During training, the agent interacts with the environment in an episodic manner, where each episode corresponds to a full 24-hour cycle of simulated time. At each control decision point, the agent selects the duration of the active green phase while respecting the fixed phase sequence and clearance intervals defined in the signal program.

To ensure realistic signal operation, several constraints are imposed during training and evaluation. Minimum and maximum green times are set to 15~s and 60~s, respectively, to prevent excessively short or long green phases that are impractical in real-world deployments. Clearance (yellow) intervals are fixed at 3~s, are identical across all control strategies, and are not subject to learning. The phase order remains unchanged throughout all experiments, isolating the learning task to green-time allocation only. All vehicles are modeled as homogeneous passenger cars, and no future traffic demand information is available to the controller at execution time.

The proposed learning-based controller is evaluated against two conventional baselines. The first is a fixed-time signal control strategy, which follows a predefined timing plan with constant green durations that do not adapt to real-time traffic conditions. Fixed-time control is widely used in practice due to its simplicity and ease of deployment.

The second is a vehicle-actuated controller, which adapts green-time allocation to real-time queue observations and represents the current state of practice in adaptive signal timing. The actuated controller allocates green time proportionally to the observed queue length on the currently served phase approaches, following 

\begin{equation}
g_t = G_{\min} + \min\!\left(\frac{Q_t}{T},\,1\right)\left(G_{\max} - G_{\min}\right),
\end{equation}

where $Q_t$ denotes the total number of halting vehicles on the active green phase's approach lanes. $T$ is a saturation threshold, and $G_{\min} = 15$~s and $G_{\max} = 60$~s are the same timing constraints imposed on the PPO policy. The saturation threshold $T = 50~\text{vehicles}$ was selected via a grid search over the full 24-hour evaluation horizon, yielding the lowest average vehicle delay while maintaining genuinely queue-responsive behavior. Including both a non-adaptive (fixed-time) and a rule-based adaptive (actuated) baseline enables a clear assessment of the extent to which learning-based green-time allocation improves performance over both conventional operation and conventional adaptation.

Performance is evaluated using standard traffic efficiency metrics. Average vehicle delay (s/veh) is computed as the total vehicle-seconds of halting (defined as time at a speed below 0.1~m/s) accumulated across all simulation steps and divided by the total number of vehicle departures over the evaluation horizon. Average queue length (veh) is the time-averaged number of halted vehicles across all controlled lanes. Throughput (veh) is the total number of vehicle departures over the evaluation horizon. In addition, environmental performance is assessed using SUMO's built-in HBEFA-based emission model; total CO\textsubscript{2} emissions and fuel consumption are accumulated over the evaluation horizon and reported on a per-vehicle basis, enabling relative comparison of the environmental impact of each control strategy. These indicators are selected to capture complementary dimensions of intersection performance: average vehicle delay is the primary user-experienced cost and the standard measure of effectiveness for signalized intersections; queue length captures spatial congestion and the risk of spillback into upstream links; throughput verifies that delay reductions are not obtained by suppressing served demand; and emissions and fuel consumption quantify the environmental dimension of increasing importance to municipal traffic management. For fair comparison, all control strategies are evaluated under identical traffic demand profiles and simulation settings, and all metrics are computed identically across strategies. All experiments are repeated across five independent random seeds, and results are reported as the mean and standard deviation across seeds to assess statistical reliability.

\subsection{Experiment~1: Baseline Comparison under Medium Traffic Demand}

The first experiment evaluates the performance of the proposed PPO-based controller under a representative medium traffic demand scenario. This demand level corresponds to typical weekday traffic conditions and serves as the primary operating point for assessing overall effectiveness.

The PPO agent is trained and tested on traffic demand drawn from the same medium-volume profile. This experiment establishes a reference performance level under nominal operating conditions and quantifies the benefits of adaptive green-time allocation when traffic demand is neither undersaturated nor heavily congested.

\subsection{Experiment~2: Robustness to Demand Perturbations}

Real-world traffic data are inherently noisy and subject to measurement uncertainty. To evaluate robustness under such conditions, the second experiment tests the sensitivity of the learned policy to perturbations in traffic demand.

In this experiment, the PPO agent is trained under a nominal medium-demand scenario and evaluated under perturbed demand profiles, where traffic volumes are scaled by $\pm 15\%$. This test targets the first of two principal axes along which realized traffic deviates from training assumptions: magnitude error, in which the overall volume differs from the assumed demand due to traffic-count inaccuracies or day-to-day fluctuation while the temporal shape of the demand profile is preserved; the complementary axis, structural change in the demand pattern itself, is examined in Experiment~3. The $\pm 15\%$ scaling represents a moderate but nontrivial deviation of this kind while keeping the intersection within a serviceable operating regime. The objective is to assess whether the controller maintains stable and effective performance when actual traffic conditions deviate in magnitude from those observed during training.

\subsection{Experiment~3: Generalization from Weekday to Weekend Traffic Patterns}

This experiment evaluates the generalization capability of the proposed reinforcement learning--based traffic signal controller across different traffic demand distributions corresponding to weekdays and weekends. Unlike Experiment~2, which focuses on robustness to moderate demand perturbations, this experiment examines whether a policy trained on weekday traffic can transfer effectively to weekend conditions without retraining. It targets the second of the two axes introduced in Experiment~2: structural change in the temporal pattern of demand, rather than error in its magnitude. The weekday-to-weekend transition is the largest naturally occurring pattern shift observable in the available sensor data, making it the most demanding generalization test the measured demand supports.

The PPO agent is trained exclusively using traffic demand profiles derived from multiple weekdays, capturing typical commuter-driven traffic patterns characterized by pronounced peak periods and relatively predictable temporal structure. The trained policy is then evaluated on weekend traffic demand profiles, which exhibit distinct characteristics such as reduced peak intensity, shifted peak times, and more irregular arrival patterns.

By testing the learned controller on unseen weekend traffic data, this experiment assesses the extent to which the policy captures general traffic dynamics rather than overfitting to specific weekday demand profiles. Successful performance under weekend conditions would indicate that the learned green-time allocation strategy generalizes across days with fundamentally different traffic behaviors, supporting its applicability in real-world deployments where traffic patterns vary across the week.

\subsection{Experiment~4: Reward Function Ablation}

The final experiment examines the role of individual reward function components in shaping the behavior and performance of the proposed reinforcement learning–based traffic signal controller. Starting from the full reward formulation described in Section~\ref{sec:methodology}, a set of simplified reward variants is constructed by selectively removing congestion-related terms associated with vehicle waiting time and queue length.

Four reward formulations are considered: (i) the full reward including both waiting-time and queue-length penalties, (ii) a variant without the waiting-time penalty, (iii) a variant without the queue-length penalty, and (iv) a variant without both congestion-related penalties. Separate PPO agents are trained using each reward formulation under identical training conditions to isolate the impact of reward design on the learned control policy.

All reward variants are trained using the average weekday traffic profile and evaluated over a full 24-hour horizon under three demand conditions: nominal demand, increased demand (+15\%), and decreased demand (-15\%). Performance is assessed using average vehicle delay and average queue length, which capture temporal delay and spatial congestion, respectively. Metrics are aggregated across the three demand conditions to enable a fair and robust comparison between reward formulations.

This ablation study complements the preceding experiments by explicitly assessing the sensitivity of the proposed controller to reward design choices. Together with Experiments~1–3, these results provide a comprehensive evaluation of the controller’s performance under nominal operation, demand perturbations, varying traffic patterns, and alternative reward formulations.

\section{RESULTS AND DISCUSSION}
\label{sec:results}

\subsection{Experiment~1: Baseline Performance Under Medium Traffic Demand}
\label{subsec:exp1_baseline}

\begin{figure*}[t]
\centering
\includegraphics[width=0.95\textwidth]{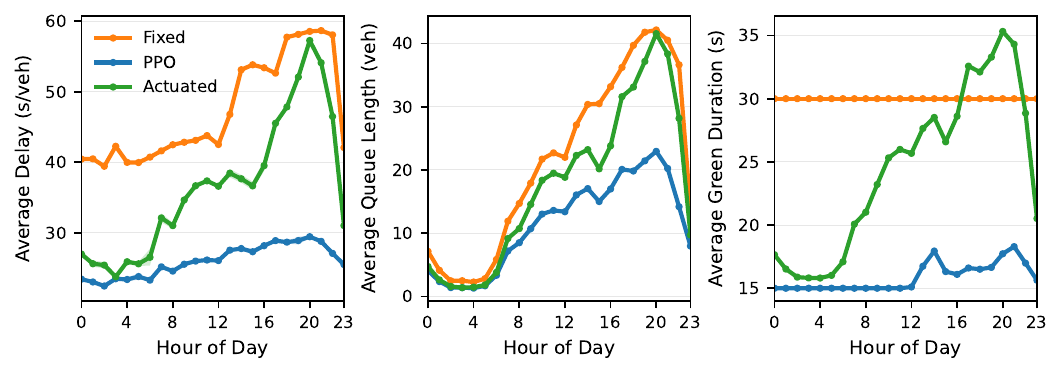}
\caption{Experiment~1 (medium traffic demand): hourly-aggregated comparison of the proposed PPO-based controller against fixed-time and vehicle-actuated control, averaged over five seeds. From left to right: average vehicle delay, average queue length, and the adaptive green-phase duration selected by each controller. Shaded bands denote $\pm1$ standard deviation across seeds.}
\label{fig:exp1_combined}
\end{figure*}

\begin{table*}[t]
\centering
\caption{Baseline performance comparison under nominal medium demand (mean $\pm$ std over five seeds). Lower is better for all metrics except throughput.}
\label{tab:baseline}
\footnotesize
\setlength{\tabcolsep}{6pt}
\begin{tabular}{lccccc}
\hline
Controller & Delay (s/veh) & Queue (veh) & Throughput (veh) & CO\textsubscript{2} (g/veh) & Fuel (g/veh) \\
\hline
Fixed-time & 50.69\,$\pm$\,0.07 & 21.25\,$\pm$\,0.03 & 36223\,$\pm$\,0 & 261.2\,$\pm$\,0.2 & 83.3\,$\pm$\,0.1 \\
Actuated & 41.52\,$\pm$\,0.07 & 17.41\,$\pm$\,0.03 & 36227\,$\pm$\,4 & 236.3\,$\pm$\,0.1 & 75.4\,$\pm$\,0.04 \\
PPO & \textbf{27.22\,$\pm$\,0.07} & \textbf{11.41\,$\pm$\,0.03} & 36228\,$\pm$\,3 & \textbf{200.4\,$\pm$\,0.2} & \textbf{63.9\,$\pm$\,0.07} \\
\hline
\end{tabular}
\end{table*}

This experiment evaluates the performance of the proposed reinforcement learning-based traffic signal controller under representative medium traffic demand conditions and compares it against both the conventional fixed-time strategy and the vehicle-actuated baseline. The objective is to assess whether adaptive green-time allocation learned via reinforcement learning can improve traffic efficiency under nominal operating conditions relative to both non-adaptive and rule-based adaptive control.

Table~\ref{tab:baseline} summarizes the results, averaged over five seeds. The proposed PPO controller achieves an average vehicle delay of $27.22$~s/veh, compared to $41.52$~s/veh for the actuated controller and $50.69$~s/veh for fixed-time control. This corresponds to a delay reduction of $46.3\%$ relative to fixed-time control and $34.4\%$ relative to the actuated baseline. The standard deviation across seeds is below $0.5\%$ of the mean for every controller and metric, indicating highly stable and reproducible performance.

The left panel of Fig.~\ref{fig:exp1_combined} reports the average vehicle delay aggregated on an hourly basis. The PPO controller maintains the lowest delay throughout the day, and its advantage widens during the peak demand periods: while the delay of the fixed-time and actuated controllers rises sharply in the evening peak, the learned policy sustains substantially lower delay, indicating that it allocates green time more effectively precisely when congestion is most severe.

The corresponding average queue length results are shown in the middle panel of Fig.~\ref{fig:exp1_combined}. Relative to fixed-time operation, the proposed controller reduces the time-aggregated queue length by $46.3\%$, and by $34.4\%$ relative to the actuated controller. This reduction reflects more efficient vehicle discharge and improved utilization of available green time under adaptive control, particularly during periods of sustained demand.

By contrast, throughput is nearly identical across all three controllers (approximately $36{,}225$ vehicles over the 24-hour horizon, differing by fewer than five vehicles). This reflects the capacity-constrained nature of the isolated intersection: under the same demand and geometry, all controllers ultimately serve the same total volume, so the performance differences manifest in how efficiently that volume is served---through reduced delay and shorter queues---rather than in raw throughput.

To provide insight into the learned control behavior, the right panel of Fig.~\ref{fig:exp1_combined} illustrates the average green-phase duration selected by each controller. Unlike fixed-time control, which applies a constant green duration, the PPO policy adaptively varies green times over the day, holding green near the minimum during low-demand hours and extending it during peak periods. This responsive allocation, distinct from the rule-based pattern of the actuated controller, explains the observed reductions in delay and queue length.

Beyond operational efficiency, the controllers differ substantially in environmental impact. As shown in Table~\ref{tab:baseline}, the PPO controller produces approximately $200$~g of CO\textsubscript{2} per vehicle, compared to $236$~g for the actuated controller and $261$~g for fixed-time control, corresponding to relative reductions of $23.3\%$ and $15.2\%$, respectively. Fuel consumption follows the same ordering and the same relative reductions. These gains are a direct consequence of reduced idling and smoother progression: lower delay translates into less time spent stationary with engines running. As the emission values are derived from a simulation-based model, they are reported as relative comparisons between controllers rather than absolute field measurements, consistent with the simulation-to-reality considerations discussed in Section~\ref{sec:methodology}.

Overall, Experiment~1 demonstrates that reinforcement learning-based green-time control significantly outperforms both fixed-time and vehicle-actuated signal operation under medium traffic demand, while also reducing emissions. These findings validate the effectiveness of the proposed approach under nominal conditions and establish a strong baseline for the subsequent robustness and generalization experiments.

\subsection{Experiment~2: Robustness to Demand Perturbations}
\label{subsec:exp2_robustness}

This experiment evaluates the robustness of the learned PPO-based controller to demand uncertainty by perturbing the nominal traffic demand profile by $\pm15\%$ while keeping the trained policy unchanged. The objective is to assess whether performance gains persist when the realized traffic volume deviates from the conditions used during training, which is representative of practical traffic-count errors and day-to-day fluctuations.

\begin{figure}[!t]
\centering
\includegraphics[width=0.95\linewidth]{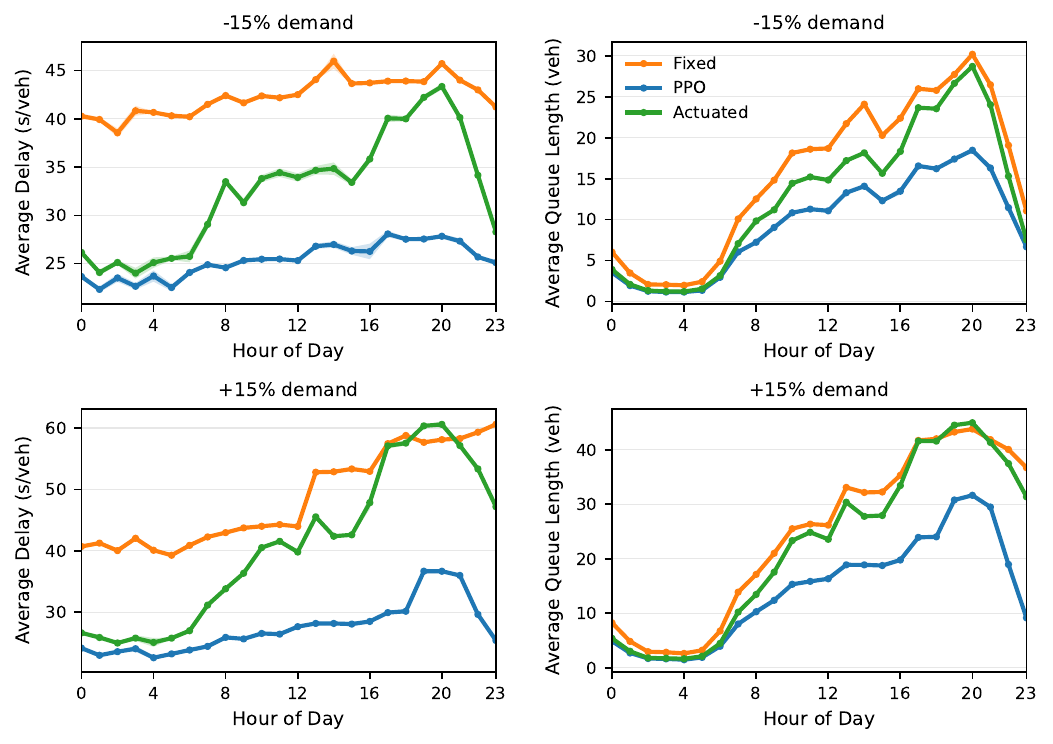}
\caption{Experiment~2 (robustness): performance under demand perturbations of $-15\%$ and $+15\%$ relative to the nominal profile (hourly aggregation, averaged over five seeds). The left column shows average vehicle delay and the right column average queue length, with the top and bottom rows corresponding to the $-15\%$ and $+15\%$ demand scenarios, respectively. All panels compare the proposed PPO-based controller against fixed-time and vehicle-actuated control. Shaded bands denote $\pm1$ standard deviation across seeds.}
\label{fig:exp2_robustness}
\end{figure}

As shown in Fig.~\ref{fig:exp2_robustness}, the proposed controller maintains a clear advantage over both baselines under each perturbation. Under the $-15\%$ demand scenario, the PPO policy achieves an average vehicle delay of $26.31$~s/veh, a reduction of $39.3\%$ relative to fixed-time control ($43.34$~s/veh) and $26.4\%$ relative to the actuated controller ($35.75$~s/veh). Under the $+15\%$ demand scenario, where congestion is more severe, the advantage is even larger: the PPO controller attains $29.43$~s/veh, reducing delay by $43.5\%$ relative to fixed-time control ($52.11$~s/veh) and $37.1\%$ relative to the actuated controller ($46.78$~s/veh). Notably, under the heavier $+15\%$ load the actuated controller's performance degrades toward that of fixed-time control during peak hours, whereas the learned policy sustains substantially lower delay, indicating that its advantage is most pronounced precisely when the network is most stressed.

Consistent trends are observed for queue length in the right column of Fig.~\ref{fig:exp2_robustness}. Relative to fixed-time operation, the PPO-based controller reduces the time-aggregated queue length by $39.3\%$ under the $-15\%$ scenario and by $41.6\%$ under the $+15\%$ scenario. As in the baseline experiment, the standard deviation across seeds remains below $0.5\%$ of the mean at both demand levels, confirming that the robustness of the learned policy is itself statistically reliable rather than an artifact of a particular random seed. These results indicate that the learned adaptive green-time allocation policy remains effective under moderate demand shifts in both directions, demonstrating robustness to realistic mismatches between expected and realized traffic volumes.

\FloatBarrier

\subsection{Experiment~3: Generalization from Weekday to Weekend Traffic Patterns}
\label{subsec:exp3_generalization}

This experiment evaluates the ability of the learned PPO-based traffic signal controller to generalize across days with fundamentally different traffic demand characteristics. The controller is trained exclusively on weekday traffic profiles and evaluated on unseen weekend demand profiles, which exhibit reduced commuter traffic, shifted peak periods, and more irregular arrival patterns.

\begin{figure}[!t]
\centering
\includegraphics[width=0.95\linewidth]{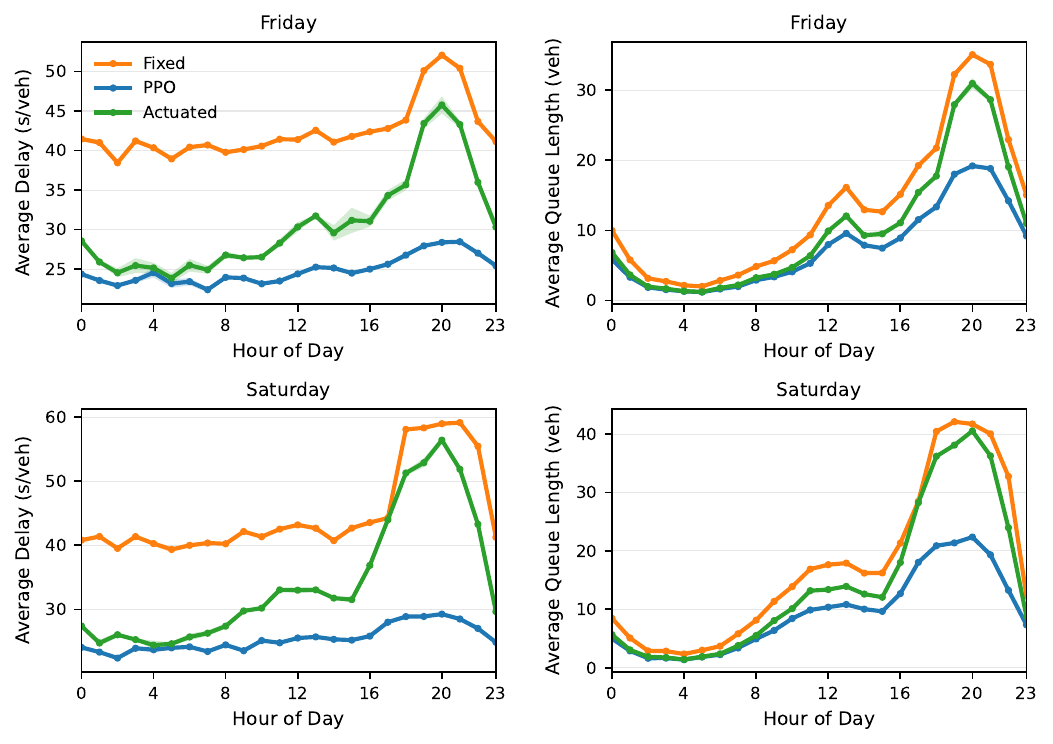}
\caption{Experiment~3 (generalization): performance of a weekday-trained PPO controller evaluated on unseen weekend traffic profiles (hourly aggregation, averaged over five seeds). Panels compare the proposed controller against fixed-time and vehicle-actuated control for the two weekend days, Friday and Saturday; left column shows average vehicle delay and right column shows average queue length. Shaded bands denote $\pm1$ standard deviation across seeds.}
\label{fig:exp3_generalization}
\end{figure}

Fig.~\ref{fig:exp3_generalization} presents the average vehicle delay and queue length for the two weekend days, Friday and Saturday. Despite the distributional shift from weekday to weekend traffic, the PPO-based controller consistently outperforms both baselines. On Friday, the learned policy achieves an average vehicle delay of $25.94$~s/veh, a reduction of $41.7\%$ relative to fixed-time control ($44.47$~s/veh) and $25.2\%$ relative to the actuated controller ($34.67$~s/veh). On Saturday, the policy achieves $26.61$~s/veh, reducing delay by $45.2\%$ relative to fixed-time control ($48.56$~s/veh) and $33.8\%$ relative to the actuated controller ($40.19$~s/veh). This result indicates that the controller captures general traffic dynamics rather than overfitting to specific weekday demand patterns.

Consistent performance gains are observed for queue length. Compared to fixed-time control, the PPO-based controller reduces the time-aggregated queue length by $41.7\%$ on Friday and $45.2\%$ on Saturday, with comparable reductions relative to the actuated controller. As in the preceding experiments, the standard deviation across seeds remains below $0.5\%$ of the mean, confirming that this cross-day generalization is statistically reliable. These reductions demonstrate that the learned green-time allocation strategy remains effective under qualitatively different traffic conditions without requiring retraining or explicit knowledge of the day type.

Overall, the results of Experiment~3 confirm that the proposed reinforcement learning--based controller generalizes effectively from weekday to weekend traffic patterns. Notably, the magnitude of improvement observed under weekend traffic conditions is consistent with the baseline performance gains reported in Experiment~1, indicating that the learned policy maintains similar efficiency benefits even when evaluated on unseen traffic patterns. This cross-day transfer capability is essential for real-world deployment, where traffic demand varies systematically across the week and retraining for each day type is impractical.

\subsection{Experiment~4: Reward Function Ablation}
\label{subsec:exp4}

This experiment investigates the contribution of individual reward components to the performance of the proposed reinforcement learning--based traffic signal controller. The baseline reward function incorporates congestion-aware penalties designed to reduce both vehicle waiting time and queue accumulation. To assess the necessity of these components, three ablated reward variants were trained by selectively removing congestion-related terms while keeping all other training conditions identical.

Four reward formulations were evaluated:
\begin{itemize}
    \item \textbf{Full reward}: includes both waiting-time and queue-length penalties.
    \item \textbf{No waiting}: removes the waiting-time penalty.
    \item \textbf{No queue}: removes the queue-length penalty.
    \item \textbf{No waiting + no queue}: removes both congestion-related penalties.
\end{itemize}

All ablation models were trained using the average weekday traffic profile and evaluated under three demand conditions: nominal demand, increased demand ($+15\%$), and decreased demand ($-15\%$). Performance metrics were aggregated across these three scenarios to ensure a fair and robust comparison. The primary evaluation metrics are average vehicle delay (s/veh), computed as the total vehicle-seconds of halting divided by total vehicle departures, and average queue length (veh), computed as the time-averaged number of queued vehicles over the full 24-hour simulation horizon. For reference, the fixed-time and vehicle-actuated baselines are included under identical conditions.

\begin{figure}[!t]
\centering
\includegraphics[width=0.49\linewidth]{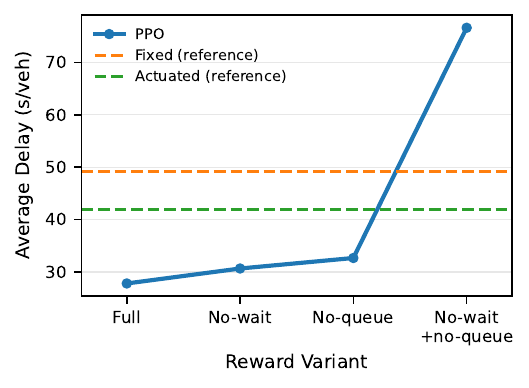}
\includegraphics[width=0.49\linewidth]{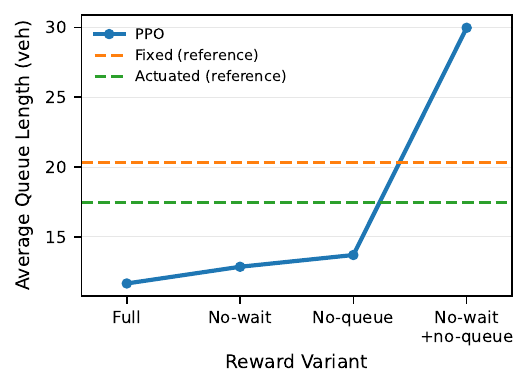}
\caption{Experiment~4 (reward ablation): average vehicle delay (left) and queue length (right) for the full reward and three ablated variants, with the fixed-time and actuated baselines shown as reference lines. Performance degrades as congestion terms are removed; with both removed, the policy falls below both baselines.}
\label{fig:exp4_ablation}
\end{figure}

Table~\ref{tab:reward_ablation} summarizes the aggregated results, and Fig.~\ref{fig:exp4_ablation} visualizes them against the two baselines. Removing either congestion-related term leads to a measurable degradation in performance. Eliminating the waiting-time penalty increases average vehicle delay by $10.3\%$ and average queue length by a comparable margin. Removing the queue-length penalty results in a larger degradation, with delay increasing by $17.5\%$ and queue length by a similar amount. Crucially, even with a single penalty removed, the controller still outperforms both the fixed-time and actuated baselines, indicating that the learned policy degrades gracefully rather than catastrophically when partially deprived of congestion feedback.

In contrast, removing both congestion penalties causes a severe collapse in control performance: average vehicle delay rises by $175.5\%$ (to $76.61$~s/veh) and average queue length by $157.2\%$. In this regime the controller performs markedly worse than both conventional baselines---its delay exceeds that of fixed-time control ($49.11$~s/veh) by more than $50\%$---confirming that without any congestion signal the agent has no useful basis for green-time allocation. This crossover, where the fully ablated policy falls below both reference baselines, is clearly visible in Fig.~\ref{fig:exp4_ablation}.

These results demonstrate that the waiting-time and queue-length penalties play complementary roles in regulating temporal delay and spatial congestion, respectively. While the queue-length term contributes somewhat more to performance than the waiting-time term, incorporating both is essential for achieving the stable, robust control demonstrated in the preceding experiments.

\begin{table}[t]
\centering
\caption{Reward ablation results aggregated over nominal and perturbed demand conditions ($\pm15\%$), mean over five seeds. Fixed-time and actuated baselines are shown for reference. Lower values indicate better performance.}
\label{tab:reward_ablation}
\footnotesize
\setlength{\tabcolsep}{3pt}
\begin{tabular}{lcc}
\hline
Reward variant & Avg.\ delay (s/veh) $\downarrow$ & Avg.\ queue (veh) $\downarrow$ \\
\hline
Full reward & \textbf{27.81} & \textbf{11.66} \\
No waiting & 30.67 (+10.3\%) & 12.86 (+10.3\%) \\
No queue & 32.67 (+17.5\%) & 13.70 (+17.5\%) \\
No waiting + no queue & 76.61 (+175.5\%) & 29.98 (+157.2\%) \\
\hline
Fixed-time (reference) & 49.11 & 20.34 \\
Actuated (reference) & 41.88 & 17.49 \\
\hline
\end{tabular}
\end{table}

\section{CONCLUSION}
\label{sec:conclusion}

This paper presented a reinforcement learning--based traffic signal control framework for adaptive green-time allocation at a single urban intersection. Using a PPO-based controller trained and evaluated in a realistic simulation environment with real-world traffic demand profiles, the proposed approach was shown to significantly outperform both conventional fixed-time control and a vehicle-actuated controller representing the current state of practice, across multiple evaluation scenarios.

Experimental results demonstrate that the learned controller achieves substantial reductions in average vehicle delay and queue length under nominal operating conditions, reducing delay by $46.3\%$ relative to fixed-time control and $34.4\%$ relative to actuated control. Robustness experiments further show that these performance gains persist under moderate demand perturbations of $\pm15\%$, indicating resilience to realistic traffic-count inaccuracies, with the controller's advantage being most pronounced under heavier demand. Generalization experiments confirm that a policy trained on weekday traffic patterns transfers effectively to unseen weekend conditions, achieving improvement levels consistent with the baseline experiment. Reward function ablation results further indicate that the proposed reward formulation plays a critical role in achieving stable and consistent performance gains by jointly accounting for vehicle waiting time and queue length; removing both congestion terms degrades performance below that of the conventional baselines. The controller additionally reduces per-vehicle CO\textsubscript{2} emissions and fuel consumption by approximately $23\%$ relative to fixed-time control, a direct consequence of reduced idling. Across all experiments, the low variance over five random seeds confirms that these results are statistically reliable. Collectively, these findings suggest that the proposed controller captures fundamental traffic dynamics related to both temporal delay and spatial congestion, rather than overfitting to specific demand profiles.

From a practical perspective, the proposed approach operates within realistic signal timing constraints and does not rely on future traffic demand information, making it suitable for real-world deployment scenarios. Because it draws on aggregate measurements from fixed infrastructure sensors rather than connected-vehicle data, its operation is also independent of connected-vehicle penetration. The demonstrated robustness and cross-day generalization are particularly important for urban environments where traffic patterns vary across time and are subject to uncertainty.

Despite these promising results, this study is limited to a single isolated intersection and a simulation-based evaluation. While this setting enables controlled experimentation and systematic analysis, it does not capture network-level interactions among adjacent intersections. In particular, optimizing an isolated intersection may redistribute congestion to neighboring nodes rather than resolving it network-wide, as improved discharge at one junction can increase arrival pressure at the next. Addressing these network-level effects requires coordinated control across multiple intersections and represents an important direction for future work.

Future research will extend the proposed framework to coordinated multi-intersection control, investigating scalable learning architectures and communication mechanisms that enable corridor and network-level optimization. These extensions will explicitly account for inter-section dependencies and bottleneck propagation while preserving the robustness and generalization capabilities demonstrated here. A further direction is the integration of connected-vehicle data through vehicle-to-infrastructure communication, extending the present infrastructure-sensing approach toward fully connected operation and enabling study of communication-layer effects such as latency and packet loss.

Overall, this work establishes a strong foundation for reinforcement learning--based traffic signal control and provides empirical evidence of its effectiveness, robustness, and generalization in realistic urban traffic settings.

\section*{ACKNOWLEDGMENT}
This work was supported and funded by Kuwait University Research Grant No. [ZE01/26]. The traffic data used in this study were obtained from the General Traffic Department of the Ministry of Interior, Kuwait. Due to confidentiality and data-sharing restrictions imposed by the data provider, the datasets are not publicly available. Access to the data may be granted upon reasonable request and subject to approval by the Ministry of Interior. AI-based language assistance tools were used solely for grammar and stylistic editing. The author reviewed and approved all content and assumes full responsibility for the work.

\bibliographystyle{IEEEtran}
\bibliography{bib}

\begin{IEEEbiography}[{\includegraphics[width=1in,height=1.25in,clip,keepaspectratio]{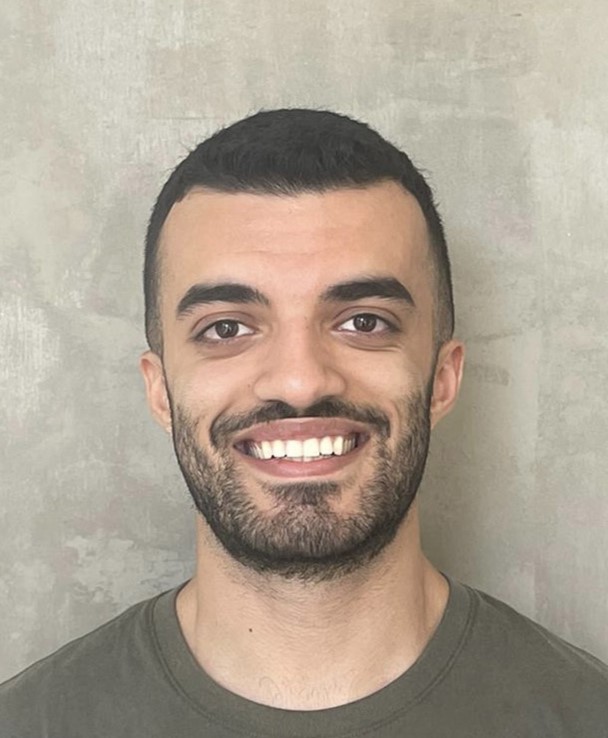}}]{Yousef AlSaqabi } (Member, IEEE) is an Assistant Professor at Kuwait University. He received his PhD in Electrical Engineering from the University of Southern California in 2024. He received both his BS and MS degrees in Electrical Engineering from The Pennsylvania State University in 2018. His research focuses on the privacy and security of vehicular networks, as well as the application of reinforcement learning and data-driven intelligence for Internet of Vehicles (IoV), connected and autonomous vehicles, and smart-city transportation systems. 
\end{IEEEbiography}

\end{document}